\documentclass[a4paper]{jpconf}
\usepackage{graphicx,amsmath,amssymb,url}

\newcommand{\td}[3]{\frac{d^{#3} #1}{d {#2}^{#3}}} 
\renewcommand{\v}[1]{\ensuremath{\mathbf{#1}}} 

\renewcommand{\bar}[1]{\ensuremath{\overline{#1}}}

\begin{document}
\title{What Can Gamma-rays from Space tell us About the Madala Hypothesis?}

\author{Geoff Beck \& Sergio Colafrancesco }

\address{School of Physics, University of the Witwatersrand, Private Bag 3, WITS-2050, Johannesburg, South Africa}

\ead{geoffrey.beck@wits.ac.za, sergio.colafrancesco@wits.ac.za}

\begin{abstract}
The recent Madala hypothesis, a conjecture that seeks to explain anomalies within Large Hadron Collider (LHC) data (particularly in the transverse momentum of the Higgs boson), is interesting for more than just a statistical hint at unknown and unpredicted physics. This is because the model itself contains additional new particles that may serve as Dark Matter (DM) candidates. These particles interact with the Standard Model via a scalar mediator boson $S$. More interesting still, the conjectured mass range for the DM candidate ($65$ - $100$ GeV) lies within the region of models viable to try explain the recent Galactic Centre (GC) gamma-ray excess seen by Fermi Large Area Telescope (Fermi-LAT) and the High Energy Stereoscopic System (HESS). Therefore, assuming $S$ decays promptly, it should be possible to check what constraints are imposed upon the effective DM annihilation cross-section in the Madala scenario by hunting signatures of $S$ decay that follows DM annihilation within dense astrophysical structures. In order to make use of existing data, we use the Reticulum II dwarf galaxy and the galactic centre gamma-ray excess data sets from Fermi-LAT, and compare these to the consequences of various decay paths for $S$ in the aforementioned environments. We find that, based on this existing data, we can limit $\tau$ lepton, quark, direct gamma-ray, and weak boson channels to levels below the canonical relic cross-section. This allows us to set new limits on the branching ratios of $S$ decay, which can rule out a Higgs-like decay branching for $S$, in the case where the Madala DM candidate is assumed to comprise all DM.
\end{abstract}

\section{Introduction}
The unknown nature of dark matter (DM) sits as a major eye-sore on the completeness of the concordance cosmology. Multiple methods are being pursued in order to catch a glimpse of this elusive substance. In particular, collider searches like at the LHC ATLAS~\cite{atlas-docs} and CMS~\cite{cms-docs} experiments can probe numerous models of theoretical interest~\cite{lhc1,lhc2,lhc3}. However, there is one promising candidate to be found within the Madala hypothesis~\cite{madala1,madala2}. This hypothesis was put forward to explain anomalies in existing LHC data but also conveniently provides a DM candidate (see Section~\ref{sec:madala} for details). This candidate is especially interesting because kinematic considerations place it in the mass range $65$ to $100$ GeV~\cite{madala2}. This region  is significant as it lies within collection of models that might be compatible with explaining the gamma-ray excess seen in the galactic centre~\cite{hessexcess,fermiexcess} by Fermi-LAT~\cite{fermi-docs} and HESS~\cite{hess-details}, as well as anti-particle excesses, according to the consideration of modelling uncertainties performed by \cite{calore2014}.

For these reasons we will examine the consequences of the annihilation of DM introduced by the Madala hypothesis within both the galactic centre and the promising dwarf galaxy target Reticulum II. Using Fermi-LAT gamma-ray data from these targets~\cite{Fermidwarves2015,daylan2016} we can formulate limits on possible annihilation cross-sections for multiple channels connecting the Madala DM to the Standard Model by comparing the existing data to gamma-ray fluxes expected from DM annihilation within these astrophysical structures. In addition to this we will examine the scalar $S$, that mediates with the DM particle in the Madala scenario, comparing what limits we can place on the decay branching ratios of $S$ to the assumption $S$ is Higgs-like~\cite{madala2}. This will be performed under the assumption that Madala DM constitutes all DM in the universe (and thus has an annihilation cross-section set to the canonical relic value~\cite{jungman1996}).

This paper is structured as follows: in Section~\ref{sec:gc} we provide more detail on the galactic centre gamma-ray excess as well as observations of Reticulum II. In Section~\ref{sec:madala} we will detail the relevant aspects of the Madala hypothesis and its bearing on the problem of DM. In Section~\ref{sec:theory} we will detail the formalism used to calculate gamma-ray emmisions from GC and Reticulum II. Finally, in Section~\ref{sec:end} we present and discuss our results.

\section{The Galactic Centre Gamma-ray Excess and Reticulum  II}
\label{sec:gc}

The galactic centre has been a prime target for DM hunts in gamma-rays ever since the discovery of unexpected gamma-ray excesses around 1 - 10 GeV~\cite{hessexcess,fermiexcess}. This is evinced by the numerous works dedicated to using DM to explain the excess~\cite{dmgc1,dmgc2,dmgc3,dmgc4,dmgc5,dmgc6,dmgc7}. In our work here we will make use of the spectrum for the excess within the region of interest (ROI) between 1$^{\circ}$ and 20$^{\circ}$ from the galactic centre found by \cite{daylan2016}. This choice of ROI is a standard when analysing galactic centre gamma-ray data (see \cite{daylan2016} and references therein) in order to avoid the emissions of the powerful Sagittarius A$^*$ complex within the galactic core region.

Reticulum II is a faint dwarf galaxy recently found by the Dark Energy Survey (DES) project~\cite{des,desdwarf}. It is notable because it is calculated to posses a very large J-factor~\cite{bonnivard2015}, a parameter that tracks the density of dark matter within halos. The density is abetted by the fact that Reticulum II is very close to Earth (30 kpc away~\cite{desdwarf}), thus reducing inverse-square flux attenuation. This particular dwarf was the source of some speculation, as there appeared a small excess in its gamma-ray spectrum observed by Fermi-LAT~\cite{geringer-sameth2015} (attributed by the aforementioned authors at $2\sigma$ confidence level to a WIMP with relic cross-section and around $60$ GeV mass). However, subsequent Fermi collaboration analysis revealed that the excess could not associated with DM when considered against other dwarf galaxy targets~\cite{Fermidwarves2015}. This was reinforced by an analysis in \cite{beck2016}, which concluded that the Reticulum II DM model would produce unacceptable excesses in the radio and gamma-ray spectra of other targets. Despite this, Reticulum II makes an ideal test-bed for producing constraints on DM models, as a large J-factor means large DM-induced fluxes from annihilation, which can then be compared to the Fermi-LAT upper limits. We make use of the upper limits established on gamma-ray fluxes from Reticulum II as used in \cite{Fermidwarves2015}.

\section{Madala Hypothesis and Dark Matter}
\label{sec:madala}

\begin{figure}[htbp]
\centering
\includegraphics[scale=0.2]{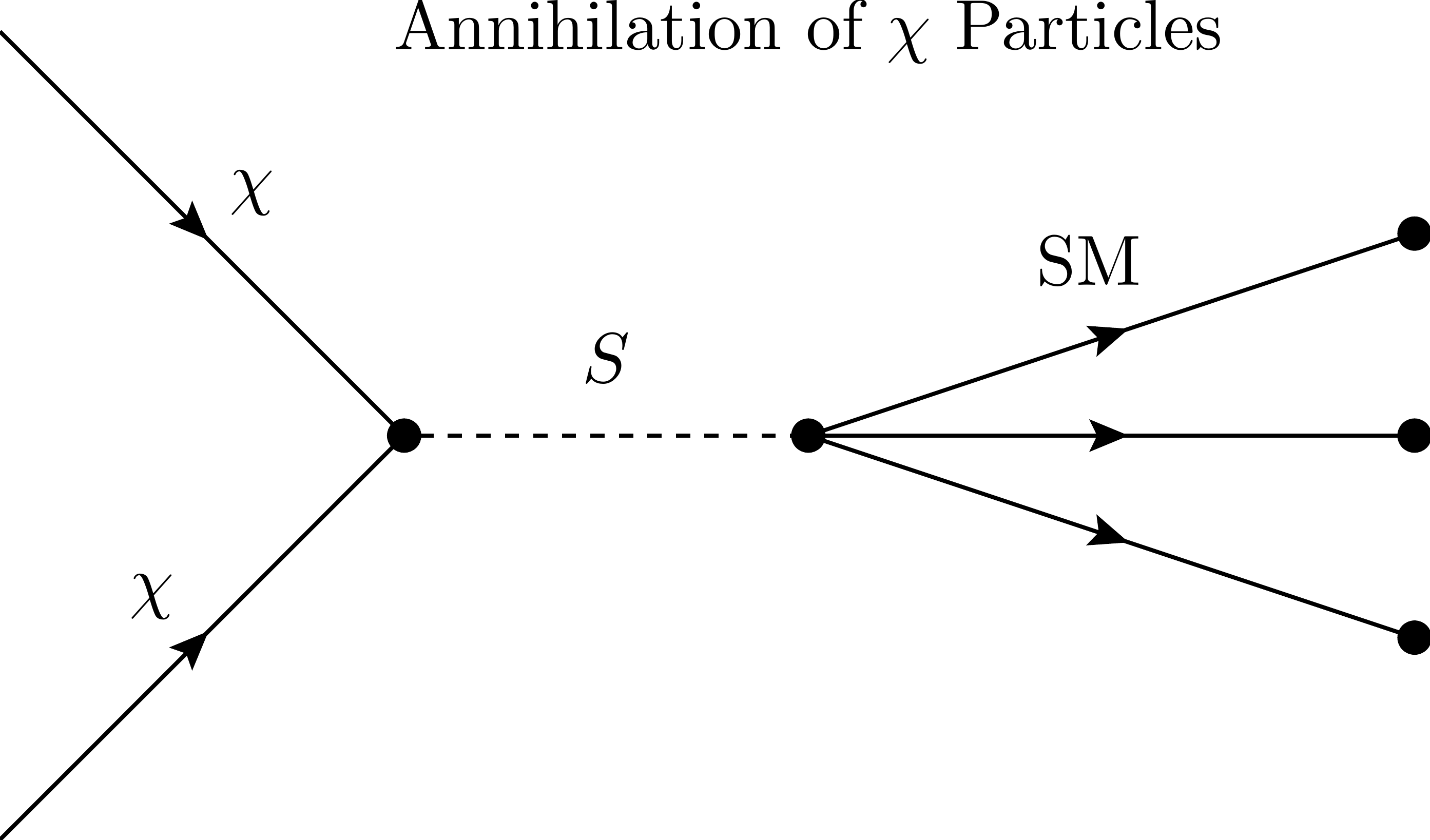}
\caption{A diagram of $\chi$s annihilating into $S$ which promptly decays to some SM particles.}
\label{fig:feyn}
\end{figure}

The Madala hypothesis is one that sets out to explain anomalies in LHC data concerning the transverse momentum of the Higgs boson within LHC collisions (among others - see \cite{madala1,madala2}). It does so through the introduction of a set of particles: a large scalar ``Madala" boson $H$ ($\sim 270$ GeV mass) which is Higgs-like, and a scalar $S$ with a mass range $130$ - $200$ GeV which couples a dark particle $\chi$ to the standard model and to $H$. This scalar $S$ acts as mediator between the standard model and dark particles proposed in the Madala hypothesis (as seen in Fig.~\ref{fig:feyn}). Therefore, in order to explore what kinds of indirect astrophysical signatures might be expected from DM that results from this conjecture, one must concentrate upon the possible couplings, and thus decay paths, between $S$ and the Standard Model. Such limits may be of particular interest in determining which couplings to the standard model are permissible for $S$ given current astrophysical data.

\section{Dark Matter halos and gamma-ray flux}
\label{sec:theory}

In a given DM halo, the differential gamma-ray flux resulting from annihilation can be specified by
\begin{equation}
\phi (E_{\gamma}, \Delta \Omega, l) = \frac{1}{4\pi} \frac{\langle \sigma V\rangle}{2 m_{\chi}^2} \td{N_{\gamma}}{E_{\gamma}}{} \int_{\Delta \Omega}\int_{l} \rho^2 (\v{r}) dl^{\prime}d\Omega^{\prime} \; ,\label{eq:diff}
\end{equation}
where $E_{\gamma}$ is the gamma-ray energy, $m_{\chi}$ is the mass of the WIMP, $\rho$ is the DM halo spatial density profile, $\langle \sigma V\rangle$ is the velocity averaged thermal annihilation cross-section, and $\td{N_{\gamma}}{E_{\gamma}}{}$ is the $\gamma$-ray yield from $S$ decay following WIMP annihilations (sourced from PYTHIA~\cite{pythia} routines in DarkSUSY~\cite{darkSUSY} as well as \cite{ppdmcb1,ppdmcb2}). In this work we will assume $S$ decays promptly, thus we will place a limit on the effective annihilation cross-section for the process $\chi\chi \rightarrow S \rightarrow SM$.

The expression Eq.~(\ref{eq:diff}) can be simplified by splitting it into two factors. The first is the astrophysical ``J"-factor, which encompasses the above two integrals,
\begin{equation}
J (\Delta \Omega, l) = \int_{\Delta \Omega}\int_{l} \rho^2 (\v{r}) dl^{\prime}d\Omega^{\prime} \; , \label{eq:jfactor}
\end{equation}
with the integral being extended over the line of sight $l$, and $\Delta \Omega$ is the observed solid angle.\\
The second factor is determined only by particle physics:
\begin{equation}
\psi (E_{\gamma}) = \frac{1}{4\pi} \frac{\langle \sigma V\rangle}{2 m_{\chi}^2} \td{N_{\gamma}}{E_{\gamma}}{} \; .
\end{equation}
Thus the flux will be found from
\begin{equation}
\phi (E_{\gamma}) =  \psi (E_{\gamma}) \times J(\Delta \Omega, l) \; .
\end{equation}

For the Reticulum II dwarf we take the J-factor to be $2.0 \times 10^{19}$ GeV$^2$ cm$^{-5}$~\cite{bonnivard2015}. However, for the galactic centre we follow the methodology of \cite{daylan2016} and use a contracted NFW profile~\cite{nfw1996}
\begin{equation}
\begin{aligned}
\rho_N(r,r_s,\eta)=\frac{\rho_s}{\left(\frac{r}{r_s}\right)^{\eta}\left(1+\frac{r}{r_s}\right)^{3-\eta}} \; ,
\end{aligned}
\label{eq:nfw}
\end{equation}
where we take the scale radius $r_s = 20$ kpc following~\cite{daylan2016}, and $\rho_s$ is defined by ensuring that $\rho = 0.4$ GeV cm$^{-3}$ at a radius of 8.5 kpc. We then calculate the average J-factor for a profile with $\eta = 1.2$ for the ROI between 1$^{\circ}$ and 20$^{\circ}$ (as explained in Section~\ref{sec:gc}) from the galactic centre using formula Eq.~(\ref{eq:jfactor}). We choose this particular profile as contracted NFW halos are often found in best-fit scenarios for DM explanations of the GC excess~\cite{daylan2016,calore2014,dmgc4}, thus making these results more relevant to the existing literature.

\section{Results and Discussion}
\label{sec:end}

\begin{figure}[htbp]
\centering
\includegraphics[scale=0.45]{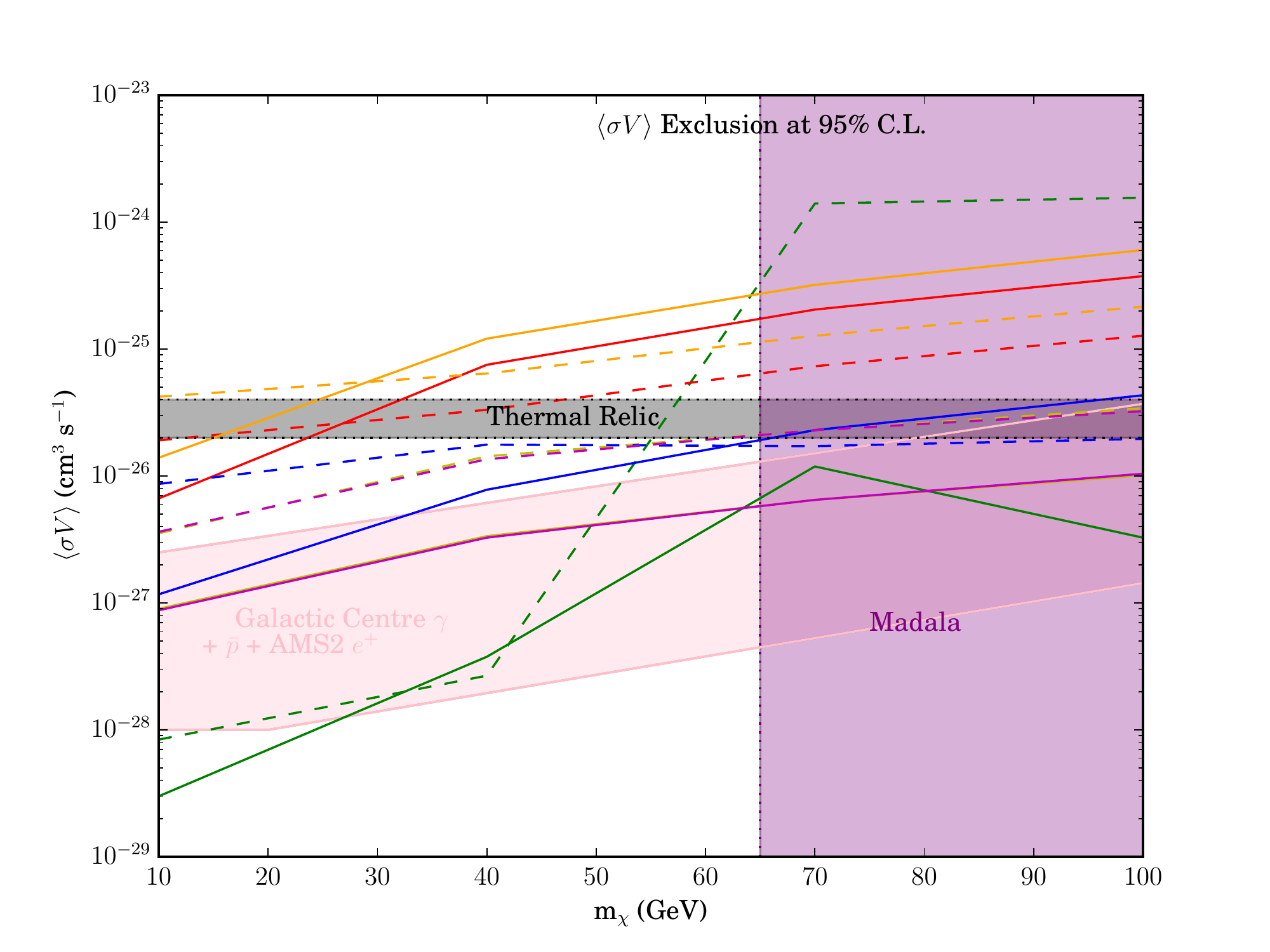}
\includegraphics[scale=0.45]{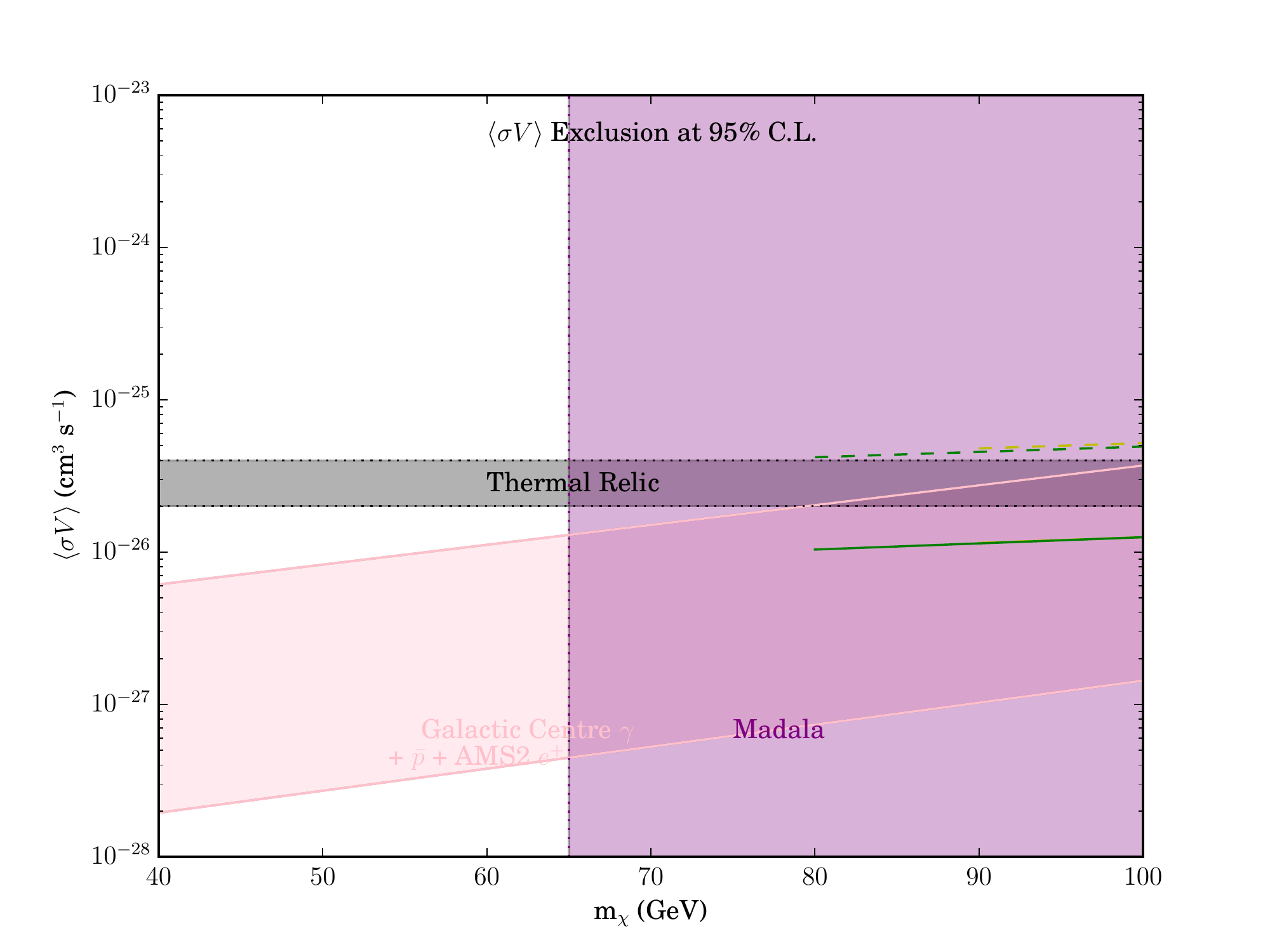}
\caption{Constraints on effective annihilation cross-section as a function of WIMP mass for various channels shown in comparison to galactic centre excess model region (pink). Reticulum II  constraints are shown as solid lines while galactic centre results are shown as dashed lines. Top: $q\bar{q}$ in yellow, Higgs in green, $e^+e^-$ in red, $\tau^-\tau^-$ in blue, $\mu^+\mu^-$ in orange, and direct gamma-rays in magenta. Bottom: $W^+W^-$ in green and $Z$ bosons in yellow.}
\label{fig:sigv}
\end{figure}

In Figure~\ref{fig:sigv} we display the limits derived from gamma-ray fluxes in the galactic centre and Reticulum II on the effective annihilation cross-section into a variety of standard model particles (assuming a branching fraction of $b_f = 1$ for each channel individually). The most significant region of the plot is the purple Madala mass band, where the DM masses correspond with the mass range expected for the boson $S$. Here we see that, for the channels $\tau^+\tau^-$, $q\bar{q}$, and direct gamma-ray production, we can use both targets to explore the region of the parameter space in which the Madala DM constitutes all the DM in the universe (relic band). In the case of Reticulum II we can also do this for the Higgs channel. For the weak boson channels we can probe below the relic band of cross-sections only in Reticulum II. We note that neither of these targets can be used to rule out the entire region of the Madala band that overlaps the galactic centre excess region from~\cite{calore2014}. However, we stress that Reticulum II has only extant upper-limits on its gamma-ray flux, thus we might expect these already strong constraints to improve with further observations.

The significance of these limits can be understood as follows: if the cross-section can be constrained below the relic level then we rule out the DM model as a candidate for all DM, as its present abundance would be too great to match cosmological constraints. However, since we assume $b_f = 1$ in each case, we can instead derive a limit on the decay branching for $S$ should the Madala DM particle constitute all DM (assuming $\langle \sigma V \rangle = 3 \times 10^{-26}$ cm$^3$ s$^{-1}$ for all channels). The results of this analysis are shown in Table~\ref{tab:summary}. Any entry with a dash signifies that no constraint can be derived (we leave out the light leptons as this is true for all masses). The final column shows the branching ratios for the Standard Model Higgs boson with mass equal to $S$ in each case (as $S$ is assumed Higgs-like~\cite{madala2}).

What is evident from these results is that both decay of $S$ into quarks and into gamma-rays as well as $W$ and $Z$ bosons (not shown on table) can be constrained below $\sim 30\%$. For $W$ bosons this means that the branching of $S$ rules out the Higgs-like case (which ranges from $\sim 0.5$ to $\sim 0.75$ in the suggested $S$ mass range~\cite{smhiggs}). While for $q\bar{q}$ our constraint rules out the Higgs-like range for $S$ masses below $\lesssim 145$ GeV. Our constraints for direct photon and $Z$ decays cannot rule out a Higgs-like $S$. Limits on decay of $S$ into Higgs bosons show large variability with WIMP mass. This arises from the movement of the gamma-ray resonant peak within the data domain for each source. Due to the hardness of the resulting gamma-ray spectrum~\cite{beck2016}, the $\tau$ lepton channel is subject to weaker constraints, with $\sim 70\%$ limit being possible for the largest masses considered, thus this channel does not affect the Higgs-like $S$ case. The branching ratios of $S$ into light leptons cannot be constrained at all.

This leaves considerable room for a Madala WIMP to constitute all DM, but raises serious doubts about a Higgs-like $S$ if the Madala hypothesis is to account for $> 50\%$ of DM. Finally, future radio frequency searches with the Square Kilometre Array~\cite{ska} can probe significantly lower cross-sections~\cite{beck2016} (even in the presence of large radio background fluxes) and will thus be integral in further constraint, or dismissal, of the Madala hypothesis and its attendant particles.

\begin{table}[htbp]
\centering
\begin{tabular}{|l|l|l|l|l|l|l|}
\hline
Channel & m$_{\chi}$ & $\langle \sigma V \rangle_f$ (GC) & $b_f$ (GC) & $\langle \sigma V \rangle_f$ (Ret. II) & $b_f$ (Ret. II) & $b_f$~\cite{smhiggs} \\
\hline
$q\bar{q}$ & 40 & 1.49e-26 & 0.5 & 3.37e-27 & 0.11 & 0.87\\
$q\bar{q}$ & 70 & 2.38e-26 & 0.79 & 6.46e-27 & 0.22 & 0.33\\
$q\bar{q}$ & 100 & - & - & 1.03e-26 & 0.34 & 2.5e-3\\
\hline
$h$ & 40 & 3e-28 & 0.01 & 3.79e-28 & 0.013 & - \\
$h$ & 70 & - & - & 1.24e-26 & 0.41 & - \\
$h$ & 100 & - & - & 3.22e-27 & 0.11 & - \\
\hline
$\tau^+\tau^-$ & 40 & 1.8e-26 & 0.6 & 7.79e-27 & 0.26 & 8.3e-2 \\
$\tau^+\tau^-$ & 70 & 1.72e-26 & 0.57 & 2.27e-26 & 0.76 & 3.5e-2 \\
$\tau^+\tau^-$ & 100 & 1.97e-26 & 0.66 & - & - & 2.9e-4 \\
\hline
$\gamma$ & 40 & 1.36e-26 & 0.45 & 3.37e-27 & 0.11 & 9.2e-4\\
$\gamma$ & 70 & 2.27e-26 & 0.76 & 6.46e-27 & 0.22 & 1.9e-3\\
$\gamma$ & 100 & - & - & 1.03e-26 & 0.34 & 5.5e-5\\
\hline
\end{tabular}
\caption{Table showing constraints on Madala DM for various masses and channels derived from the galactic centre (GC) and Reticulum II (Ret. II). Both a constraint for $\langle \sigma V \rangle$ assuming $b_f = 1$ is shown as well as a constraint on the decay branching when it is assumed $\langle \sigma V \rangle$ takes the value of the relic cross-section. This $b_f$ can be compared against the Standard Model Higgs results for the appropriate mass choice~\cite{smhiggs}.}
\label{tab:summary}
\end{table}

\section*{Acknowledgments}
S.C. acknowledges support by the South African Research Chairs Initiative of the Department of Science and Technology and National Research Foundation and by the Square Kilometre Array (SKA). G.B acknowledges support from a post-doctoral grant through the same initiative and institutions.

\end{document}